%% file: main.tex
\DocumentMetadata{}
\documentclass[sigconf]{acmart}

\makeatletter
\def\@ACM@checkaffil{
    \if@ACM@instpresent\else
    \ClassWarningNoLine{\@classname}{No institution present for an affiliation}%
    \fi
    \if@ACM@countrypresent\else
        \ClassWarningNoLine{\@classname}{No country present for an affiliation}%
    \fi
} 
\makeatother
\usepackage{acmart-taps}

\AtBeginDocument{%
    \providecommand\BibTeX{{%
        \normalfont B\kern-0.5em{\scshape i\kern-0.25em b}\kern-0.8em\TeX}
    }
}

\def\sysname{Proscenium}
\usepackage{xspace}
\usepackage{balance}
\usepackage{color}
\usepackage{bm}

\def\etal{\textit{et~al.}\xspace}
\def\eg{\textit{e.g.,}\xspace}
\def\incl{\textit{incl.}\xspace}

\copyrightyear{2026}
\acmYear{2026}
\setcopyright{cc}
\setcctype{by}
\acmConference[CHI EA '26]{Extended Abstracts of the 2026 CHI Conference on Human Factors in Computing Systems}{April 13--17, 2026}{Barcelona, Spain}
\acmBooktitle{Extended Abstracts of the 2026 CHI Conference on Human Factors in Computing Systems (CHI EA '26), April 13--17, 2026, Barcelona, Spain}
\acmDOI{10.1145/3772363.3798304}
\acmISBN{979-8-4007-2281-3/2026/04}

\begin{document}

\title[\sysname]{\sysname: Exploring Design Spaces of Layered Information Experience on a Large Dual-Layer Transparent Display}

\author{Chen Chen}
\orcid{0000-0001-7179-0861}
\email{chechen@fiu.edu}
\affiliation{%
  \institution{Florida International University}
  \city{Miami}
  \state{FL}
  \country{USA}
}
\affiliation{
  \institution{Microsoft Research}
  \city{Redmond}
  \state{WA}
  \country{USA}
}

\author{Michel Pahud}
\orcid{0000-0003-3924-1189}
\email{mpahud@microsoft.com}
\affiliation{%
  \institution{Microsoft Research}
  \city{Redmond}
  \state{WA}
  \country{USA}
}

\author{David Brown}
\orcid{0000-0002-6013-2713}
\email{dave.brown@microsoft.com}
\affiliation{%
  \institution{Microsoft Research}
  \city{Redmond}
  \state{WA}
  \country{USA}
}

\author{Chuck Needham}
\orcid{0000-0002-0088-4681}
\email{chuck@chuckneedham.com}
\affiliation{%
  \institution{Microsoft Research}
  \city{Redmond}
  \state{WA}
  \country{USA}
}

\author{Balasaravanan T. Kumaravel}
\orcid{0000-0002-4816-2063}
\email{bala@tkbala.com}
\affiliation{
  \institution{Microsoft Research}
  \city{Redmond}
  \state{WA}
  \country{USA}
}

\author{Andrew D. Wilson}
\orcid{0009-0000-5369-6863}
\email{awilson@microsoft.com}
\affiliation{%
  \institution{Microsoft Research}
  \city{Redmond}
  \state{WA}
  \country{USA}
}

\author{Ken Hinckley}
\orcid{0000-0002-4733-4927}
\email{kenneth.p.hinckley@gmail.com}
\affiliation{%
  \institution{Microsoft Research}
  \city{Redmond}
  \state{WA}
  \country{USA}
}

\author{Nicolai Marquardt}
\orcid{0000-0002-5473-2448}
\email{nicmarquardt@microsoft.com}
\affiliation{%
  \institution{Microsoft Research}
  \city{Redmond}
  \state{WA}
  \country{USA}
}

\renewcommand{\shortauthors}{Chen~\etal}

\input{0-abstract}



\begin{teaserfigure}
    \centering
    \includegraphics[width=\linewidth]{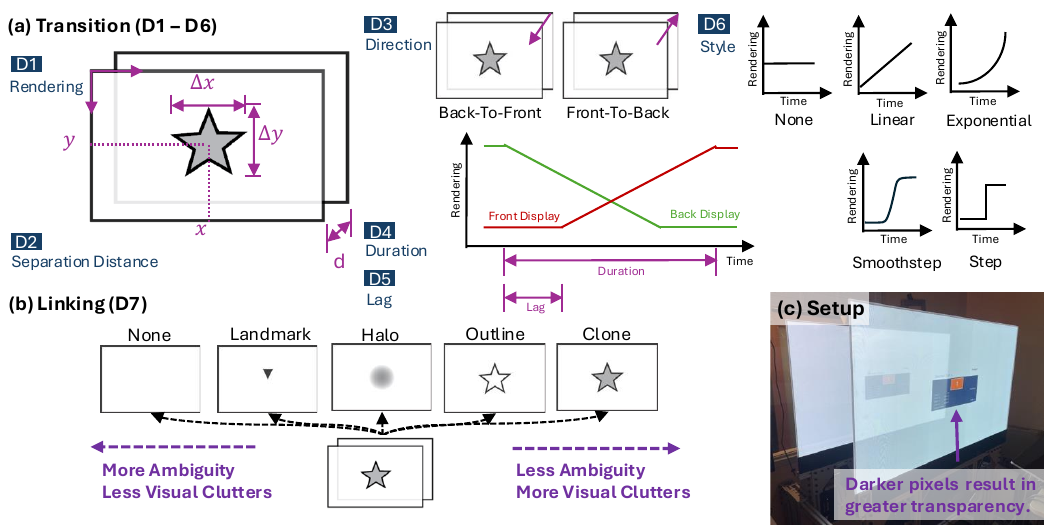}
    \caption{Design space of \sysname~workstation in terms of \emph{transition} (a) and \emph{linking} (b). (c) Setup of \sysname ~workstation.}
    \Description{Subplots (a) and (b) demonstrate the design space of the Proscenium workstation in terms of ``transition'' and ``linking''. The category of ``transition'' encompasses six dimensions, including ``rendering'', ``direction'', ``style'', ``separation distance'', ``duration'', and ``lag''. The category of ``linking'' encompasses five types, including ``none'', ``landmark'', ``halo'', ``outline'', and ``clone''. Subplot (c) shows the setup of the Proscenium display, with two parallel, transparent displays.}
    \label{fig::designspace}
\end{teaserfigure}

\maketitle

\input{1-introduction}

\input{2-design-space}

\input{3-tool}
\input{4-app}

\input{5-conclusion}

\input{6-ack}

\balance
\bibliographystyle{ACM-Reference-Format}
\bibliography{reference}

\appendix
\input{a01-implementation}

\end{document}

%% file: 0-abstract.tex
\begin{abstract}
Layering information spaces is a promising strategy to design intuitive and engaging interactive experiences.
Although multi-layer displays enable promising interaction techniques through limited depth perception - achieved via slight separation between layers - it remains unclear how to fully design experiences that leverage the unique affordances of layered information.
To address this, we introduce \sysname, a dual-layer, large transparent display workspace setup with an adjustable separation between the layers.
We demonstrate our preliminary design space focusing on how rendered information can be \emph{transitioned} and \emph{linked} across displays, and showcase $14$~ speculative experience prototypes across six categories.
\end{abstract}

%% file: 1-introduction.tex
\section{Introduction}\label{sec::intro}
Layering information is a critical strategy to design intuitive and engaging interaction experiences.
It allows the foreground to highlight critical information while preserving contextual awareness in the background.
It also opens up new design opportunities to dynamically transition entities between layers based on real-world contexts.
For example, smart whiteboard like Range~\cite{Ju2008} shows how the sketch content can be pushed to the background by overlaying a transparent backdrop when users' physical proximity increases.

Layered information can be realized on a traditional 2D display by simply adjusting the transparency of the foreground and background layers.
However, rendering multiple layers on single display can be unintuitive and causing visual clutter.
Another approach to distinguish layers is to use \textbf{M}ulti-\textbf{L}ayer \textbf{D}isplays (MLD) like PureDepth~\cite{PureDepth}, featuring a layered LCD monitor with a transparent display overlaying a second screen. 
Prema~\etal~\cite{Prema2006} speculated on possible methods for rendering objects across the display, but how the interaction experience could be designed to take advantage of the limited depth remains unclear.
The minimal separation between the displays may also restrict depth perception.
Similar concepts of using layered transparent displays to create optical illusions have also been applied in modern artwork, though interactive experiences have yet to be explored~\cite{Homeli, Taggart2018}.
While mixed reality headsets and spatial displays (\eg~~\cite{LookingGlasses}) can easily render layered information, their reliance on additional instruments and/or computational costs may limit their use in many applications.

To explore how we can design headset-free layered information experience, we introduce \emph{\sysname}\footnote{``\sysname''~ is originally a theater staging design metaphor that refers to the use of vertical space to emphasize the main scene~\cite{Pilbrow2016}.}, a dual-layer large transparent display workspace setup with an adjustable distance between the displays.
Rather than using a non-transparent back display (\eg~\cite{PureDepth}), layering two transparent displays enhances space utilization and allows the rendered objects to be more visually integrated into the work environment.
While the number of display layers can exceed two, where an infinite number of layers would be equivalent to a volumetric display - we only focus on a dual-layer setup - a first step toward understanding the design space of layering information and interacting with such experiences.
Future work may explore beyond-dual-layer experience with our insights.
With \sysname, we present our preliminary design space and showcase $14$~speculative experience prototypes across six categories.
Though this paper focuses on speculative design and experience prototyping, future work may validate and evaluate the proposed design space with real-world stakeholders.

%% file: 2-design-space.tex
\section{Preliminary Design Space of the Rendering Techniques}\label{sec::design}

\sysname~ presents a dual-layer transparent display setup, consisting of a front layer and a back layer, with an adjustable separation between the display (Figure~\ref{fig::designspace}c).
Figure~\ref{fig::designspace}a - b demonstrate our synthesized design space, categorized under \emph{transition} and \emph{linking}, that helps understand how experiences can be designed and interacted with \emph{across} the displays.

\vspace{4px}\noindent{\bf Transition (\fbox{D1} - \fbox{D6})}.
\emph{Transition} defines how a specific rendering such as remote users or shared content can be transitioned between the front and back layer across space (\fbox{D1} - \fbox{D3}) and time (\fbox{D4} - \fbox{D6}).
Notably, \emph{Fading} is considered as a specific type of transition, where the entity gradually appears (fading in) or disappears (fading out).

\vspace{4px}\noindent{\bf Linking (\fbox{D7})}.
Due to the visual discontinuity~\cite{Prema2006} between two displays, we used the concept of \emph{linking} to define the potential strategies to help users mentally connect two renderings on front and back layer display.
We speculated possible strategies include \emph{none}, \emph{landmark}, \emph{halo}, \emph{outline} and \emph{clone}.
Figure ~\ref{fig::designspace}b illustrates how the back layer may render information from the front layer, visually aiding viewers in connecting information across layers. 
In an extreme case, \emph{clonning} the front layer’s rendering onto the back layer can reduce ambiguity but may increase visual clutter.

%% file: 3-tool.tex
\section{\sysname~ Workstation}\label{sec::implementation}

We iteratively built a large dual-layer transparent display setup, allowing for rapid prototyping, iteration, and experimentation with a wide range of experiences and interactions.
We use two Planar LookThru LO552 55'' transparent OLED displays with full HD resolution~\cite{PlanarDisplay} (Figure ~\ref{fig::designspace}c).
Two displays were mounted in parallel on an aluminum frame with an adjustable distance between them, allowing us to vary \fbox{D2} for different application.
Rather than implementing full end-to-end applications, we use pre-recorded videos and images to enable rapid experience prototyping.

%% file: 4-app.tex
\begin{figure*}[t]
    \centering
    \includegraphics[width=\linewidth]{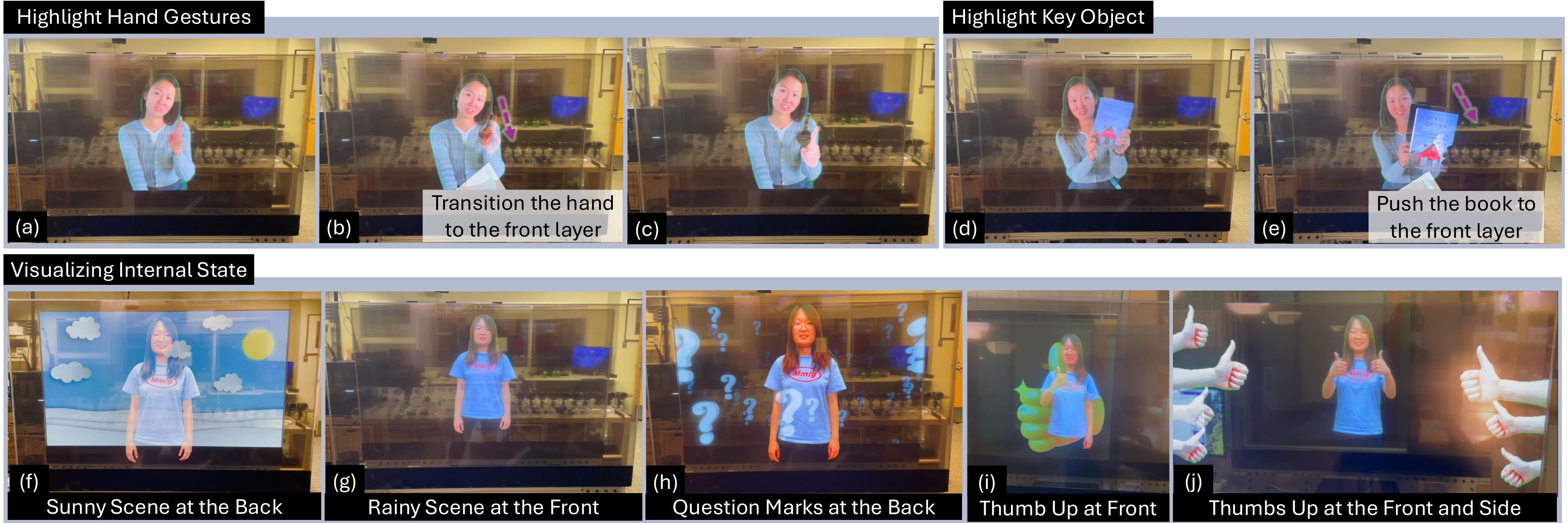}
    \caption{Experiences that highlight non-verbal cues in telepresence, \incl~emphasizing the stretch-out hand (a - c), critical object (d - e), and the internal states of the remote user (f - j).}
    \Description{Prototyped experiences that highlight non-verbal cues in telepresence, including emphasizing the stretched-out hand (demonstrated in (a) - (c)), a critical object such as a book (demonstrated in (d) - (e)), and the internal states of the remote user (demonstrated in (f) - (j)).}
    \label{fig::example_highlight}
\end{figure*}

\section{Experience Prototyping}\label{sec::app}

Grounded on the method of experience prototyping~\cite{Buchenau2000}, we present $14$ experiences that \sysname~ enables, organized across six categories.

\subsection{Highlighting Non-verbal Cues in Telepresence (\colorbox{blue!25}{E1} - \colorbox{blue!25}{E3})}\label{sec::app::highlight}

We explore how \sysname~ may be used to highlight critical non-verbal cues during remote dyadic communications.

\vspace{+4px}\noindent
{\bf \colorbox{blue!25}{E1} Emphasize the stretch-out hand using layer separation.}
We demonstrate how behaviors like stretching out critical hand gestures while the remote user is speaking can be highlighted through layer separation.
Figure ~\ref{fig::example_highlight}a - c shows an example where the remote user unconsciously extends a hand gesture for `one', while verbally saying: {\it ``first, [...]''}.
We set the separation distance (\fbox{D2}) to $72$ cm, which approximates the arm length of an adult woman~\cite{Hennessey1994}.
After experimenting with various rendering parameters (\fbox{D1}), we decided to use transparency exclusively, as it did not distort the scale of the rendered remote user.
Figure ~\ref{fig::example_highlight}b illustrates a transient state where the hand is rendered on both the front and back displays, with reduced transparency.
Once the hands transitioned to the front layer, we set the hand on the back layer to full transparency to emphasize the gestural behaviors (Figure ~\ref{fig::example_highlight}c).

\begin{figure*}[t]
    \centering
    \includegraphics[width=\linewidth]{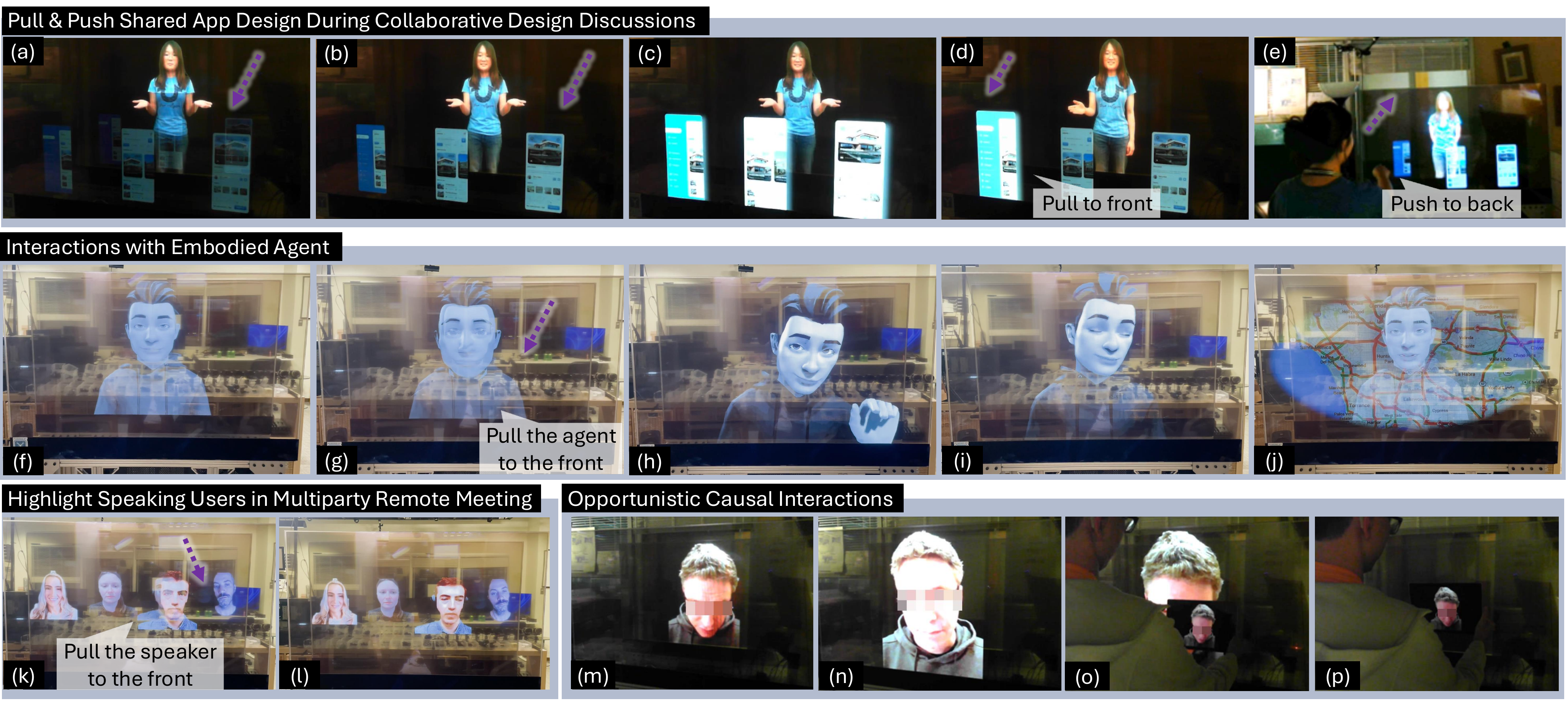}
    \caption{Experiences that demonstrate how the proximity afforded by the separation between layers can encourage the interactions, \incl~pulling and pushing the shared app design for collaborative brainstorming (a - e), interacting with an embodied agent (f - j), highlighting the speaking user in multiparty videoconferencing (k - l), and opportunistic causal interactions (m - p).}
    \Description{Experiences that demonstrate how the proximity afforded by the separation between layers can encourage the interactions. (a) - (e) demonstrate a collaborative app design with a Proscenium user and a remote meeting participant. The rendered app can be transitioned from back display to front display (i.e., pull), or from front display to back display (i.e., push). (f) - (j) shows the interaction with an embodied agent. (k) - (l) demonstrate how the talking user can be highlighted by transitioning from the back display to the front display. (m) - (p) shows the opportunistic causal interactions, in which the remote meeting participant can transition from the back display to the front display and personal tablet.}
    \label{fig::example_proxemics}
\end{figure*}

\begin{figure*}[t]
    \centering
    \includegraphics[width=\linewidth]{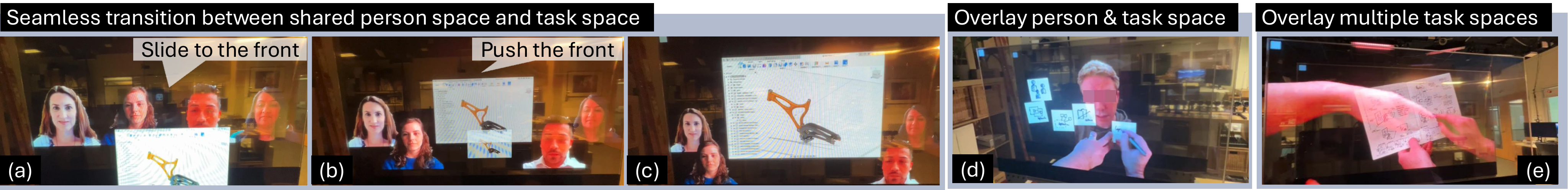}
    \caption{Experiences that provide management support of shared person and task spaces, \incl~a seamless transition between shared person space and task space (a - c), overlays person and task space (d), and multiple task spaces (e).}
    \Description{Experiences that provide management support of shared person and task spaces, incl. a seamless transition between shared person space and task space (a - c), overlays person and task space (d), and multiple task spaces (e).}
    \label{fig::example_workspace_separation}
\end{figure*}

\vspace{+4px}\noindent 
{\bf \colorbox{blue!25}{E2} Emphasize the critical key object using layer separation.}
Similar to \colorbox{blue!25}{E1}, we prototyped the experience to highlight key objects in coordination with spoken messages. 
We explored how to enhance the prominence of physical artifacts through layer separation.
\mbox{Figure ~\ref{fig::example_highlight}d - e} show how the hand-held book is extracted and gradually ``pushed'' to the front display, as the remote user is verbally emphasizing: {\it ``[...] this is a great book I read [...]''}.
We prototyped this experience using the same parameters (\fbox{D1} – \fbox{D6}) as those in \colorbox{blue!25}{E1}.
Experiencing the key objects moving \emph{physically} closer to \sysname~ users can potentially foster a more realistic, engaging and focused communication experience.

\vspace{+4px}\noindent 
{\bf \colorbox{blue!25}{E3} Visualizing internal states.}
Inspired by techniques that visualize users' internal states in 2D videoconferencing applications~\cite{ReactionOverlay, ZoomVideoAR, MSFTTeamGesture}, we demonstrate how \sysname~ can highlight the internal states in a remote videoconferencing experience. 
Figure ~\ref{fig::example_highlight}f - j showcase five representative prototyped experiences, where overlaid animations are presented and transition between the front and back displays in synchronization with the spoken message.
Specifically, Figure ~\ref{fig::example_highlight}f - h illustrate how mental states, such as happiness, sadness and confusion, can be metaphorically visualized on different display layers, drawing inspiration from real-life experiences.
\mbox{Figure ~\ref{fig::example_highlight}i - j} show how the thumbs-up and multiple thumbs-up animations can enhance remote users on the front display.

\subsection{Encouraging Interactions through Proximity \colorbox{green!25}{E4} - \colorbox{green!25}{E7}}
Our second set of experience prototyping explored how the variations of the physical proximity between rendered information and \sysname ~user can encourage (or hinder) users' intentions and engagement of the interactions.

\vspace{+4px}\noindent 
{\bf \colorbox{green!25}{E4} Pull \& push the shared app design for collaborative brainstorming.}
In a videoconferencing experience with shared task space~\cite{Buxton2009}, we explored how variations of proximity between the task and remote users, as well as task and \sysname~ users influence the communication experience.
Figure ~\ref{fig::example_proxemics}a - e illustrate how \sysname~ facilitates collaboration with a remote participant on a Figma~\cite{Figma}-like smartphone application design.
We enable the shared design to be \emph{pulled} to either the \sysname~ users (Figure ~\ref{fig::example_proxemics}d, \eg~ when the remote user is attempting to assign a new design task to the \sysname ~user) or \emph{pushed} to the remote participant (Figure ~\ref{fig::example_proxemics}e, \eg~ when one of the \sysname ~ user is attempting to emphasize the takeaways of one specific design).
This assists users in engaging in the design discussion, enabling a viewing experience when the rendered information transitioned between \emph{social} space and \emph{personal} space~\cite{hall1966hidden}.
Realizing this experience requires three layers: two layers positioned closed to and far from \sysname~ users, with one additional layer in between.
We created the visual effect of bringing information closer to \sysname~ users by slightly increasing the transparency, the size of the prototyped app design, and adding subtle shadows (Figure ~\ref{fig::example_highlight}c - d)~\cite{Zhai1996}.

\vspace{+4px}\noindent 
{\bf \colorbox{green!25}{E5} Interactions with embodied agent.}
\noindent We demonstrate how varying the proximity between the embodied agent and \sysname~ encourage more engaging interactions.
\mbox{Figure ~\ref{fig::example_proxemics}f - h} demonstrate how the life-size embodied agent transitions to the front layer, enters the personal zone of the \sysname~ user, and initiates proactive interaction such as greeting as the \sysname~ user approaches the display and enters the agent's personal space.

\vspace{+4px}\noindent 
{\bf \colorbox{green!25}{E6} Highlighting the speaking user in multiparty videoconferencing experience.}
Grounded on the ideas of proxemic interactions \cite{Marquardt2015Proxemic}, \sysname~ can enhance the multiparty videoconferencing experience by highlighting the active speaker.
Figure ~\ref{fig::example_proxemics}k - l showed how the meeting participants can be pulled and transitioned to the front display when they begin speaking.
Reducing the proximity between the speaking videoconferencing participant and the \sysname~ user, while transitioning the speaker from social space to personal space, enhances the user's awareness of the active remote participant.

\vspace{+4px}\noindent 
{\bf \colorbox{green!25}{E7} Opportunistic causal interactions.}
\sysname~ enables support for opportunistic causal interactions, which typically occur spontaneously and without prior planning.
To foster such interactions between remote users and \sysname~ users, \sysname~ would bring the remote user from the public space to social space, by transitioning the video feeds of the remote users to the front display (Figure ~\ref{fig::example_proxemics}m - n).
Once causal interactions are complete, the \sysname~ workstation enables the \sysname~ user to move the remote user to their personal space by transferring the videoconferencing to personal devices, \eg~tablet (Figure ~\ref{fig::example_proxemics}o - p).

\begin{figure*}[t]
    \centering
    \includegraphics[width=\linewidth]{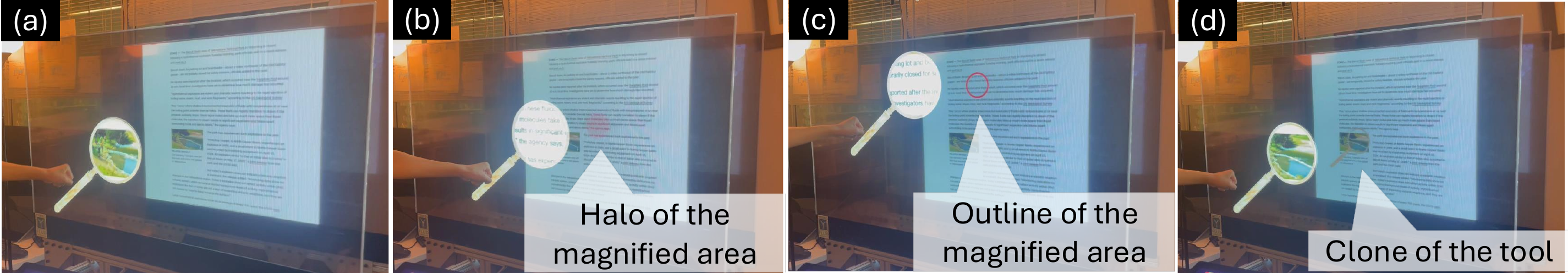}
    \caption{Experiences that demonstrate how the focused information at the front display can be \emph{linked} at the back display, using strategies of \emph{none} (a), \emph{halo} (b), \emph{outline} (c), and \emph{clone} (d).}
    \Description{Experiences that demonstrate how the focused information at the front display can be linked at the back display, using strategies of none (a), halo (b), outline (c), and clone (d).}
    \label{fig::example_tool}
\end{figure*}

\begin{figure*}[t]
    \centering
    \includegraphics[width=\linewidth]{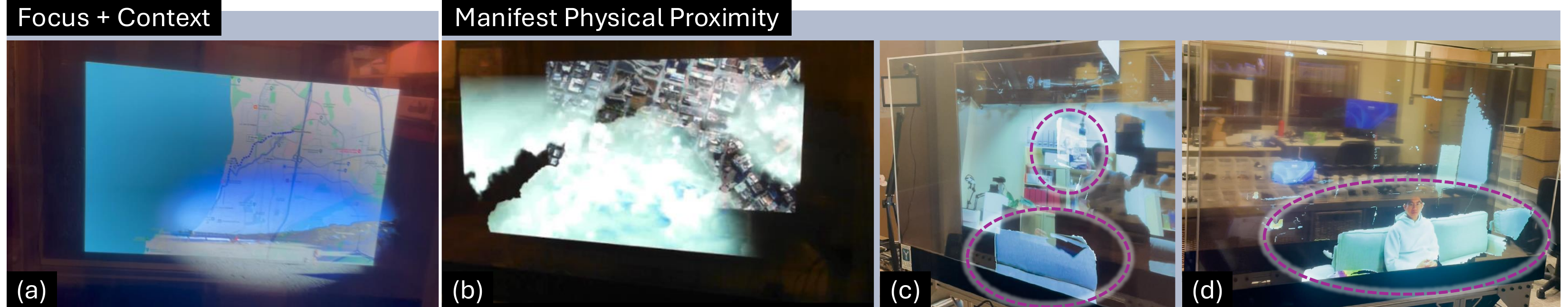}
    \caption{Experiences that demonstrate (a) overview + detail and (b - d) the manifestation of physical proximity. Key entities extracted to the front display are highlighted by a purple circle.}
    \Description{Experiences that demonstrate (a) overview + detail and (b - d) the manifestation of physical proximity. Key entities extracted to the front display, including a remote meeting participant sitting on a sofa, are highlighted by a purple circle.}
    \label{fig::focus_context_physical_proximity}
\end{figure*}

\subsection{Managing Person and Task Space (\colorbox{red!25}{E8} - \colorbox{red!25}{E10})} 
\sysname ~can support managing shared person and task spaces~\cite{Buxton1992, Buxton2009} in videoconferencing experiences.

\vspace{+4px}\noindent{\bf \colorbox{red!25}{E8}~Seamless transition between shared person space and task space.}
A \emph{seamless} transition between shared person and task spaces are critical in a videoconferencing experience~\cite{Buxton1992}.
This experience shows how the shared content can be pushed to the back display (Figure ~\ref{fig::example_workspace_separation}a - c) by adjusting center positions ($x$ and $y$) and scales ($\Delta x$ and $\Delta y$) to ensure the shared content remains fully visible.
This subtle transition allows co-located \sysname~ users to perceive how the shared workspace is presented and pushed to remote meeting participants.

\vspace{+4px}\noindent{\bf \colorbox{red!25}{E9} Overlaying person and task spaces.}
When it comes to some shared task, \sysname~ users may also need to be aware of the remote user. 
The ability to "read" remote meeting participants while maintaining focus on the task space has been recognized as a key enabler of effective telepresence~\cite{Buxton1992}.
Figure ~\ref{fig::example_workspace_separation}d demonstrates the experience when the workspace activities such as brainstorming design ideas with sticky notes are pulled to the front display while keeping remote users at the back layer.
Using physical separation to distinguish the task and user spaces creates a realistic viewing experience, describing how the task and user would be positioned if they were co-located, enabling \sysname ~ user focus on the task while being aware of the remote collaborators.

\vspace{+4px}\noindent{\bf \colorbox{red!25}{E10} Overlaying multiple task spaces.}
\sysname~ workstation can enable multiple videoconferencing participants to collaborate within a shared workspace.
We demonstrate how multiple individual workspace can be combined through overlaying.
Multiple remote participants can simultaneously interact with the shared workspace, each being rendered on a separate display layer (Figure ~\ref{fig::example_workspace_separation}e). 
This creates a visually layering experience for \sysname~ users, showing the presence of remote participants as if they were co-located.

\subsection{Separating Task and Tools (\colorbox{red!25}{E11})}

\sysname~ workstation can be used as a smart visualization tool designed for co-located meetings.
Inspired by TouchTools~\cite{Harrison2014TouchTools}, we used two distinct layers to render tools and tasks.
To assist users cognitively connect the tool with the task, we implemented four \emph{linking} strategies, \incl~ \emph{None}, \emph{Halo}, \emph{Outline}, and \emph{Clone}, shown in Figure~\ref{fig::example_tool}a - d respectively.

\subsection{Overview + Detail~(\colorbox{cyan!25}{E12}) }
%
The technique of Overview + detail ~\cite{Cockburn2009} has been widely used in map applications, presenting both the map location and detailed scene images as users navigate a space.
This experience demonstrates a design where the front display renders the scene while the back display shows the corresponding map location (Figure~\ref{fig::focus_context_physical_proximity}a).
User can gain a sense of a specific location of interest while maintaining awareness of its location on the map.

\subsection{Manifestation of Physical Proximity (\colorbox{teal!25}{E13} - \colorbox{teal!25}{E14})}\label{sec::app::meeting}

\sysname~ can arrange and render information that reflects physical proximity.

\vspace{4px}\noindent{\bf \colorbox{teal!25}{E13}~Supporting tools for examining satellite cloud maps.}
Satellite cloud maps allow visual inspection of cloud cover over specific areas, which is essential for weather forecasting, climate research, and disaster management.
With \sysname, the front display renders the clouds while the satellite view remains on the back display, demonstrating the physical proximity between cloud and ground (Figure~\ref{fig::focus_context_physical_proximity}b).

\vspace{4px}\noindent{\bf \colorbox{teal!25}{E14} 2.5D videoconferencing experiences.}
\sysname~can facilitate a $2.5$D videoconferencing experience by amplifying and emphasizing critical depth information.
Figure~\ref{fig::focus_context_physical_proximity}c illustrates how a book, hands and part of a arms can be extracted to the front display in real time, enabling \sysname~users to be aware of critical objects during communication.
This experience is similar to \colorbox{blue!25}{E1} - \colorbox{blue!25}{E2}, although we use actual proxemics to separate the information layers.
Figure~\ref{fig::focus_context_physical_proximity}d presents another example in which the scene of a remote participant sitting on a sofa is extracted to the front display, while only key objects, such as a display, are rendered on the back display.
Removing less critical background information behind remote participants can enhance the telepresence experience, making it feel as if the remote participant is physically present with the \sysname~ users.

%% file: 5-conclusion.tex
\section{Conclusion and Future Work}\label{sec::conclusions}

We introduce and explore interaction designs around \sysname~- a dual-layer, large transparent OLED display workstation.
Grounded in our speculative design space, we rapidly prototyped $14$ experiences to explore how the separation of two transparent displays can support future interactive experience design.

Our preliminary speculative design and experience prototyping point to four directions for future work: 

\emph{First}, although this paper does not include user studies (\eg~empirical evaluation, expert feedback, and/or comparisons with alternatives), future work will involve a design review with real-world stakeholders. This will allow us to iteratively refine the design and further explore the key parameters outlined in Figure~\ref{fig::designspace}a – b.

\emph{Second}, although our design space primarily focuses on transition and linking, future research on \sysname~ will also explore a broader set of design parameters.
For example, leveraging the affordances of see-through displays, future research may examine the effects of display placement (\eg~as a room divider or integrated into window fixtures).
More broadly, future work could also examine how the layered display relates to its surrounding physical environment - an exploration that a range of existing sensing technologies, \eg~\cite{Agarwal2020, Boovaraghavan2023, Chen2020CapTag}, may help enable.
Future work may also explore the impact of interaction failures and unintended visual interference.

\emph{Third}, although the experiences presented in this poster are primarily prototyped with recorded and edited video and image assets, we also aim to build functional end-to-end applications that enable evaluations in real-world, ecologically valid settings.

The \emph{final} long-term future work is to understand the impacts and designs by adding additional layers.
While this work focuses solely on the dual-layer setup as a \emph{first} step toward understanding design and interactions with layered information experiences, future research may extend beyond dual layers using our insights.
Although \sysname~is not a holographic display, future work may explore how it can support a more diverse range of layered information experiences.

%% file: 6-ack.tex
\begin{acks}

We thank the anonymous reviewers for their valuable feedback. 
We gratefully acknowledge the Project Transcendence team~\cite{Transcedence} at Microsoft Research for their support and feedback. 
We also thank Nathalie H. Riche, Kori Inkpen, John Tang, Ed Cutrell, Martez Mott, Sean Rintel, Payod Panda, Lev Tankelevitch, Hugo Romat, Gonzalo Ramos, Jialu Gao, Sneha Gathani, Jieun Kim, Xiajie Zhang, Humphrey Curtis, Zheng Ning, Anna Offenwanger, and Frederic Gmeiner from Microsoft Research for their insightful feedback during the early stages of this work.

\end{acks}

%% file: a01-implementation.tex
\section{Supplementary Details of \sysname~Workstation Setups}\label{sec::app::setups}

This section provides additional supplementary materials for Section~\ref{sec::implementation}, focusing on the setup and implementations of \sysname~ workstation.

\subsection{Hardware Setups}

\sysname~workstation uses two existing Planar LookThru LO552 55'' transparent OLED displays with full HD resolution~\cite{PlanarDisplay}.
The transparency is determined by the luminance of each pixel's color, as defined by the BT.709 standard~\cite{BT709}, where $\alpha = 0.2126r + 0.7152g + 0.0722b$.
While prior studies involving with MLD~\cite{PureDepth, Wong2003} often used LCD displays, we opted for an OLED transparent display due to its wider viewing angles, better color accuracy, and full transparency when fully black pixels are rendered.
\sysname~ workstation also includes a third display, which are used for rapid prototyping and exploring different experiences.
While primarily relying on the WOz study, where the researcher controls the rendering behavior on \sysname ~ without the users' awareness, we also integrate \sysname ~ with a hand tracking sensor and an RGB-D camera to facilitate rapid prototyping with a few experiences.

\begin{figure*}[h!]
    \centering
    \includegraphics[width=\linewidth]{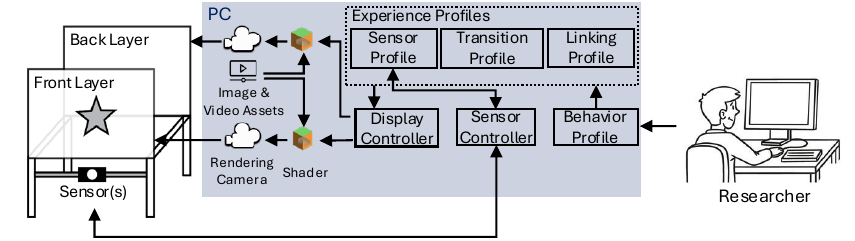}
    \caption{Hardware and software implementations of \sysname~ workstation, where designers and/or researchers can rapidly prototype a wide variety of experiences and experiment with different parameters related to transition and linking discussed in Figure~\ref{fig::designspace}.}
    \Description{Hardware and software implementations of Proscenium workstation, where designers and/or researchers can rapidly prototype a wide variety of experiences and experiment with different parameters related to transition and linking discussed in Figure 1.}
    \label{fig::software_design}
\end{figure*}

\begin{figure*}[h!]
    \centering
    \includegraphics[width=\linewidth]{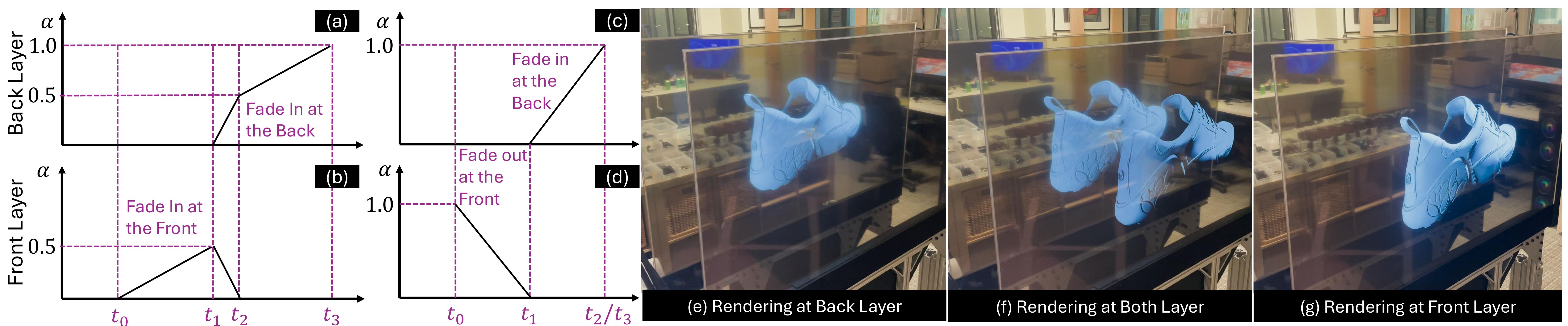}
    \caption{Example transfer function for the front and back displays resulting in transitioning and fading effect realized by adjusting transparency ($\bm{\alpha}$); (a - b) Transfer function when the information rendering is fading in at the front display, followed by back display; (c - d) Transfer function when the information rendering is transitioned from front to back display; (e - g) examples of how a rendered shoe on the back layer, both the front and back layer, and the front layer.}
    \Description{Example transfer function for the front and back displays, resulting in a transitioning and fading effect realized by adjusting transparency; (a - b) Transfer function when the information rendering is fading in at the front display, followed by back display; (c - d) Transfer function when the information rendering is transitioned from front to back display; (e - g) examples of how a rendered shoe on the back layer, both the front and back layer, and the front layer.}
    \label{fig::transition_example}
\end{figure*}

\subsection{Exploratory Tool for Supporting Rapid Experience Prototyping}\label{sec::implementation::software}

We designed and developed a rapid prototyping tool as a Windows application built with Unity.
Figure ~\ref{fig::software_design} illustrates the key components in our software design. 
During experience prototyping, our prototyping tool enables designers to rapidly and iteratively create various experience and experiment with different parameters speculated in our design space (Figure ~\ref{fig::designspace}), without requiring the development of complete and end-to-end applications~\cite{Dow2005}.
Researchers can configure profiles to define how rendered information transitions, how foreground and background information can be visually linked and how sensor stream can be used.
During the preliminary WOz study, our tool need to enable researchers to trigger key animation to simulate a virtually interactive experience.

\vspace{+4px}
\noindent {\bf Rendering information entities on dual-display setup.}
To explore the experience of \sysname~ and its speculated design space, we selected and curated workspace applications that exemplify the core concepts of \sysname's interactions (see Section~\ref{sec::app} for the in-depth discussion of the explored experience).
%
Rather than developing a complete end-to-end application from scratch, we captured a collection of images and video clips assets.
Our prototyping tool then renders the selected multimedia assets based on users' pre-configurations.
Prototyping the interactive experiences with the offline pre-recorded multimedia assets allow us to propel a rapid explorations of a wide variety of viewing experience, interactive techniques and application scenarios while enabling iteratively refining the design as we increased our prototype fidelity.
To present information on both transparent display and realize the concept of transition as well as linking, we used two Unity cameras with two customized shaders to render information on each transparent display, while an additional camera rendered the interactive GUI controller on a third display.

\vspace{+4px}
\noindent {\bf Transition across the layers.}
Transition was identified as one key dimension to be explored among various speculated experience.
Modeling and implementing the behaviors of transition using image and video assets is challenging due to diverse design dimensions that may vary across time and space.
The design of \sysname~ prototyping tool should enable researchers to rapidly prototype the style of transitioning described in Figure ~\ref{fig::designspace}.
To address this, we designed a transition curve that defines how rendering changes over time, structured around three configurable phases.
Researchers can define the transitioning style for each phase, enabling behavior prototyping for \emph{transitioning}, including \emph{fading}.
Figure ~\ref{fig::transition_example}a - b and Figure ~\ref{fig::transition_example}c - d demonstrate examples of gradually fading in the rendered shoe to the front layer and transitioning the shoe from the front layer to the back layer, respectively.
While $\alpha$ is used as the rendering parameter specified in \fbox{D1}, our tool enables researchers to experiment with other type(s) of parameters.
\mbox{Figure ~\ref{fig::transition_example}e - g} show an example of a rendered shoe on the back layer, both the front and back layers, and the front layer, respectively.

\vspace{+4px}
\noindent {\bf Instrumenting and integrating additional sensors and devices.}
Our initial design of the \sysname~ workstation features two parallel-stacked transparent displays but does not include any sensing capabilities.
Despite the possibilities of using WOz approach where the experimenter manually triggers pre-configured multimedia assets, minimal sensing capabilities may be required to prototype certain experiences that dynamically adapt the viewing experience based on changing contexts, such as the hand gestures.
The designing of our \sysname~ workstation includes two additional sensing systems:

\vspace{+4px}\noindent $\bullet$ {\bf Hand tracking.}
The transparent displays used in the prototyping of the \sysname~ workstation are limited to rendering information only and do not have touchscreen capabilities.
To track hand positions and gestures, we placed a Leap Motion camera~\cite{LeapMotion} at the bottom of the front display, continuously collecting the $27$ key joints of each hand in 3D space.
During \sysname's ~ initializations, we first compute the optimal translation and rotation matrices between the Leap Motion camera system and \sysname's prototyping tool using Kabsch-Umeyama Algorithm~\cite{Umeyama1991, Kabsch1976, Kabsch1978}.
During this calibration process, $12$ pre-set locations on the front display were used.
These transform matrices are then be stored persistently during subsequent design and study.
During the experience prototyping, our prototyping tool continuously computes the tracked hand positions within the camera system to render the pre-configured multimedia assets.

\vspace{+4px}\noindent $\bullet$ {\bf Integrations of additional tablet and mobile sensors.}
Beyond expanding \sysname~ with touchscreen functionality through hand tracking, we also integrated an additional tablet into our workstation.
Our tool enables prototyping of rendered information behaviors using the sensor stream broadcast by the tablet.
Our implementation used a Microsoft Surface Pro tablet~\cite{MSSurfacePro} and the Escapement tool~\cite{Nicholas2023Escapement} to stream sensor data.

\vspace{+4px}\noindent $\bullet$ {\bf Streaming and understanding depth-enhanced video feeds of remote users.}
\sysname~ workstation integrates a \mbox{RGB-D} camera to facilitate the exploration and experience prototyping of how remote videoconferencing participants can be rendered and presented, taking advantage of the dual transparent displays.
While most \sysname~ experiences are prototyped using pre-recorded image and video assets (Section~\ref{sec::implementation::assets}), a few are developed with higher fidelity to provide a more realistic interactive experience.
Despite the need for wireless streaming of camera videos in today's videoconferencing applications, we chose to interface a Orbec Femto Bolt camera~\cite{OrbecCam} with the \sysname~ workstation via a serial cable. 
The camera would then be placed in a separate room during exploration and experience evaluations.
While our initial prototype used the video feeds from a wireless IP camera and used recent pre-trained AI models like Depth Anything~\cite{Yang2024depth} to infer pixel-wise depth, we eventually decided on using a RGB-D camera, which greatly simplifies the development of our prototyping tool and allows us to focus on experience prototyping instead of designing network architecture for real-time high-quality video streaming.

\subsection{Assets Creations and Refinement}\label{sec::implementation::assets}
Rather than developing fully functional, end-to-end applications, experience prototyping on the \sysname~ workstation focuses on recreating the speculated experience mainly through a WOz process.
To support this process, we pre-recorded and edited a collection of image and video assets, which are subsequently rendered using the \sysname~ tool with predefined behaviors and experience profiles.
This prototyping approach propelled a rapid exploration of many techniques and experiences without the overhead of full application implementation, while allowing for iteratively design refinement.
We used widely used film techniques such as green screen~\cite{Foster2014GreenScreen} to record the footage of the users as they speaking, showing key objects and different gestural behaviors.
This enables us to efficiently remove the background while rendering users on the transparent display, creating the visual perception that remote users are colocated with \sysname~ users.
Existing AI tools like MediaPipe~\cite{Lugaresi2019MediaPipe} and Segment Anything Model~\cite{Kirillov2023SAM, Ravi2024SAM2} were used to extract hands and key objects, which would be rendered on separate display layers based on pre-defined experience profiles.